\documentclass[twocolumn,aps,floatfix,showpacs]{revtex4}
\usepackage{graphicx}
\usepackage{dcolumn}
\usepackage{bm}
\usepackage{epsf}
\usepackage{float} 
\usepackage{amssymb}
\usepackage{amsmath,amsfonts}
\newcommand{\eq}[1]{(\ref{#1})}
\newcommand{\la}{\label}
\newcommand{\bea}{\begin{eqnarray}}
\newcommand{\eea}{\end{eqnarray}}
\newcommand{\beq}{\begin{equation}}
\newcommand{\eeq}{\end{equation}}
\newcommand{\be}{\begin{equation}}
\newcommand{\ee}{\end{equation}}
\newcommand{\ii}{{\rm{i}}}
\newcommand{\dd}{{\rm{d}}}
\newcommand{\vv}{{\rm v}}

\newcommand{\p}{\partial}

\renewcommand{\>}{\rangle}
\newcommand{\<}{\langle}

\def\XXint#1#2#3{{\setbox0=\hbox{$#1{#2#3}{\int}$ }
\vcenter{\hbox{$#2#3$ }}\kern-.5\wd0}}

\begin{document}

\title{Non-Linear hydrodynamics and Fractionally Quantized Solitons at Fractional Quantum Hall Edge }

\author{P. Wiegmann}
 \affiliation{James Franck Institute, University of Chicago, 929 57th St, Chicago, IL 60637}
\date{\today}

\begin{abstract}
\noindent We argue that dynamics of  gapless  Fractional Quantum Hall Edge states is essentially non-linear and that it features  fractionally quantized solitons propagating along the  edge. Observation of  solitons would be a direct evidence of fractional charges.  We show that the non-linear dynamics of the Laughlin's  FQH state is governed by the quantum Benjamin-Ono equation. Non-linear dynamics of gapless edge states is determined by  the nature of gapped  modes in the bulk of FQH liquid. 
The dispersion of edge modes  is traced to the  double boundary layer of FQH states.
\end{abstract}  

\maketitle
\noindent\emph{1. Introduction and Results}. In a Fractional Quantum Hall state electrons  collectively  constitute an incompressible liquid almost free from dissipation. Excitations in this liquid are gapped by an energy $\Delta_\nu$ determined by the Coulomb interaction \cite{GM}.  The 
gap   is large compared with the temperature (but small
compared with the cyclotron
energy \(\Delta_\nu\ll\hbar\omega_c\)). The only low energy current carrying states in a FQH liquid are \emph{edge states}  localized on the boundary \cite{Edge}.
Edge states provide a valuable tool to probe FQH states. 

In most FQH states  the excitation gap is large compared  to the energy of long wave edge states, and for that reasons is commonly neglected. The standard approach to the theory of  Edge states starts from the  Chern-Simons action in the bulk  \cite{Wen}. This action  has no scale. It neglects gapped bulk  modes but focuses on braiding properties of FQH states.  The  Hamiltonian of this theory  differs from zero only
by a confining potential. In the case of a "sharp" boundary it  is a topological field theory where all states are in a correspondence with boundary states. They are chiral bosons with the current algebra 
 \be\label{f}[f(x),\,f^H(x')]=\frac{2\nu}{\pi }\nabla_x\delta(x-x'),\ee
 governed by a linear wave equation\begin{align}\label{w}
 \dot f-c_{0} \nabla_xf=0.  
\end{align} Here 
\be f^{H}=\frac{1}{\pi }\int \frac{%
  f(x^{\prime })-f(x)}{x'-x}dx^{\prime }\ee
   is the Hilbert transform of \(f(x)\),   \(\ell_B\) is the magnetic length $2\pi\ell_B^2=\Phi_0/B$,  $c_0=\hbar^{-1}\ell^2_B|\nabla_y
U|$ is velocity of sound determined by the slope of the confining potential
 \(U(y)\). The unperturbed droplet occupies the half plane \(y\!<\!0,\) \(x\)   is a coordinate along  the  boundary.

This theory assumes that a FQH state does not change toward the
boundary and neglects  electric polarization caused by edge waves. This happens
if  the curvature of the  potential  is small compared to the gap
 \(\ell_B^{2}\nabla^2_yU\ll\Delta_\nu\), but a slope is larger than electric field  $\ell_B^{2}\nabla_yU\gg e^2$.
Also it  assumes that boundary waves are long and their amplitudes are small compared with scales  set by the magnetic length. We  accept these conditions.

Physics  missed by this otherwise successful theory can be seen in the following setting. Let us suddenly perturb the edge by a classical instrument, say RF-source (whose spatial extent is larger than magnetic length \(\ell_B\)) and then release the system. A smooth semiclassical density profile \(f_0(x)\) will occur.  How does the density profile propagate along the edge? The wave equation (2) suggests that the initial profile simply translates as \(f(x, t) = f_0(x-c_0t)\) without changes.   This may be true   shortly after  the perturbation,   but at  time of order of \(\hbar/\Delta_\nu\) the profile  is expected to change.
This time is not too long. In typical \(\nu=1/3\) samples  it is about 10ps \cite{GM}. At that time the wave equation fails. 

Another apparent problem  of the wave equation is that  it does not discriminate between fractional and  integer QHE where a mobility
gap is due to a disorder. Do these very different cases share the same dynamics?

Corrections to  linear waves (\ref{w}) come from few sources: the curvature of the confining potential 
at the edge, mixing with higher Landau levels, disorder,  and, more interestingly, the interaction between the gapless  edge  and  gapped bulk  modes.  The latter is the most relevant for FQHE. It is the subject of
this letter.    

We show that the wave equation receives important corrections of two sorts: a nonlinear term and a dispersion term. Both are proportional to the scale \(\kappa\sim\Delta_\nu \ell_B^2/\hbar \) of gapped excitations omitted in the Chern-Simons action. We will show that they are  relevant for the non-equilibrium physics of FQHE at the edge. 

\begin{figure}%
\centering
\parbox{1.2in}{\includegraphics[scale=0.55]{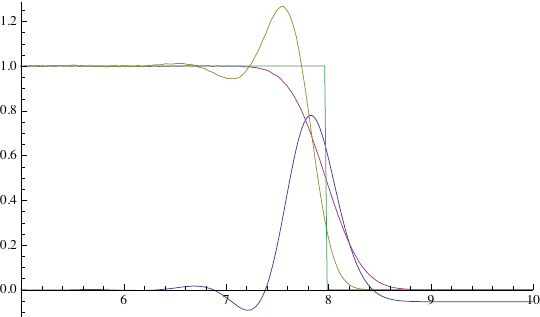}}%
\qquad
\begin{minipage}{1.2in}%
{\includegraphics[scale=0.35]{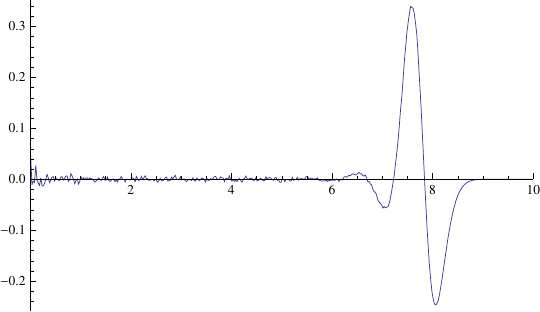}}
\end{minipage}
 \caption{\label{dipole}Boundary double layer of $\nu=\frac 13$
state: Left (in  decreasing order): density \(\rho\), 
density \(\rho_I\) at \(\nu=1\) (scaled), the dipole moment 
 \(-\eta(r)=-\int^{r}
y'(\rho-\rho_I)\dd y'\). The value of $\eta|_{\infty}=\frac{1-\nu}{4\pi}
\approx 0.05307$ is clearly
seen. Right:   $\rho-\rho_I$ illustrates Eq.\eq{A} \cite{Shytov}. The
inner boundary  lays in the bulk before
the overshoot.}
\end{figure}

In this paper we focus only on the simplest Laughlin's states with a single branch of excitations assuming that the  state remains intact all the way to the edge. In this case the quantum equation for edge modes reads
\begin{align}\la{BO}
\dot f\!-c_0\nabla_x f-\kappa \nabla_x\left( \frac{\!f^2}{2} -\eta\cdot\nabla_x f^H\right)\!=0,\; \eta=\frac{1\!-\nu}{4\pi}  
\end{align}
Here  \(\eta=\frac{1-\nu}{4\pi}\) is the
dipole moment of the \emph{boundary double layer} explained below (see \eq{d} and Fig.~\ref{dipole}).
\begin{figure}
\caption{Boundary waves: the boundary layer is highlighted}
\includegraphics[scale=0.45]{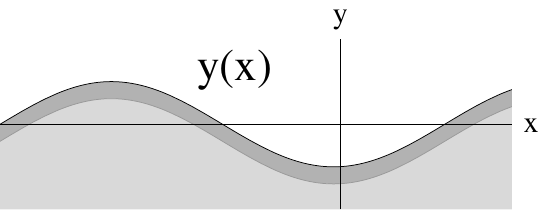}
\end{figure}

The new terms in brackets cause  new phenomena. One is 
\emph{fractionally charged solitons} on the edge.

 The chiral boson in this equation has a transparent interpretation. Acting on coherent state with an electronic density $\rho(x,y)$ the chiral boson means  a boundary density 
 \(f(x)=\int^{y(x)} _{0}\rho(x,y) dy\), where \(y(x)  \) is a boundary displacement counted from an unperturbed boundary $y=0$. The bosons act 
in the chiral Fock  space, where \(\ii\hbar\pi f^H+\nu\nabla_x\pi_f\)=0.
Here  
\(\pi_f=-\ii\hbar\frac{\delta}{\delta f}\)
 is a momentum of $f$.
 The constraint  yields the  algebra (\ref{f}).
 
The non-linear equation \eq{BO} previously appeared in two domains of physics. Its classical version  has been  derived by Benjamin \cite{BO} in 1967 for inner  waves  in a deep stratified incompressible fluid with a rapidly changing density or shear \cite{Maslowe}. It is called the Benjamin-Ono equation. The quantum Benjamin-Ono equation (qBO) identical to (\ref{w},\ref{BO}) describes the chiral sector of  Calogero model  \cite{AW,BAW4}. 

Both relations are not accidental: a FQH state is an incompressible quantum fluid with rapidly changing density and  shear at the boundary as  shown in Fig.~\ref{dipole}. Both FQH  and Calogero liquids feature excitations with a fractional charge. One can treat the result of this paper as a proof of a long anticipated equivalence between Calogero liquid and FQH Edge states.

The qBO  has an intrinsic relation to a boundary Conformal Field Theory with the central charge $c=1-6\nu^{-1}(\nu-1)^2$ situated in the exterior of the
droplet. Conformal symmetry emerges with respect to   deformations of the boundary of a FQH droplet. Also, qBO has a natural extension to non-Abelian FQH states. We will discuss these aspects elsewhere. In this paper  we mention  a few major  features of the qBO equation \cite{AW,BAW4} and focus on  its derivation.

\emph{2. Solitons in Quantum Benjamin-Ono Equation.}
The qBO  is a Hamiltonian equation. The dispersion term  \(\eta\nabla_x^2 f^H \) has the dimension of viscosity but contrary to real viscosity does not produce dissipation. It gives a non-analytic correction to the dispersion of linear waves
\be\hbar\left(\omega(k)-c_0k\right)=\kappa\eta k|k|.\ee
A noticeable feature of  the qBO is that the "dissipationless viscosity" $\eta$ is expressed
solely in terms of  the  filling fraction \(\nu\) and, in this sense, is universal. Conversely in classical liquids \cite{BO,Maslowe} the coefficient \(\eta\) depends on the equation of state. It is not quantized. Also, \(\eta\) vanishes for the  IQH. It is inherent to FQHE. It is similar in origin to the "Hall viscosity" \cite{Halviscosity}, but is not directly related.

Another noticeable feature is that qBO being a non-local equation, nevertheless is integrable \cite{ABW}.
 In integrable equations a competition between  non-linear and  dispersion terms yields \emph{solitons}. In the qBO there are two branches of solitons: one is \emph{ultrasonic}, another is \emph{subsonic}. Quite remarkably,  both carry quantized electron charges \cite{BAW4}. An ultrasonic soliton carries an integer of  electron charge \(q=+e\). It is a bump on the edge - a coherent state of an electron. A subsonic soliton is a coherent state of holes - a dent on the edge. It carries integer of a fractional charge $q=-\nu e$ of an opposite sign.
Shapes of  the elementary solitons  are especially simple:\begin{align}
f_q(x+c_0t)=\frac{q}{\pi }\frac{a}{a^2(x-v_qt)^2+1},\quad q=1,\;-\nu.
\end{align} 
Velocity of a soliton  (relative to the  sound) is  \(v_q=q\kappa\eta a\) is inversely proportional  to the magnetic field. It is proportional to its amplitude $a$ and its charge $q$. The amplitude \(a>0\)  is arbitrary, but the charge is quantized.

Notice that in the classical limit $\nu\to 0$, the fractional ("hole") branch disappears, but
the quantized  "particle" branch  remains.  BO is the unique classical equation which has quantized solitons.
Charge carried by solitons of  any other soliton equations
can be arbitrary.

When a     dent in the boundary density \(f_0(x)  \)  is created by an RF source,  corrections to the wave equation become important at time
 \(t_c\sim(\kappa|\nabla  f_0|)^{-1}\), which is only by a factor \((\ell_Bf_0)^{-1}\) larger than  the typical time \(\hbar/\Delta_\nu\sim\) 10ps. At that time a wave packet  collapses through a shock wave to oscillatory features which further separate to a stable \emph{fractionally charged  soliton train}   - a sequence of pulses with  fractionally  \(-\nu e\) charges  \cite{BAW4}. This suggests an appealing prospect for a direct observation of fractional charges: apply an RF-source to the edge, observe a sequence of fractionally charged  pulses.

\emph{3. Phenomenological Hamiltonian}  is the starting point of the analysis. But first we must identify a space where the Hamiltonian acts.  This space is the result of  projection  enforced
by the condition \(\hbar\omega_c\gg\Delta_\nu\). It is the set of states  obtained by a deformation of the Laughlin ground state  $\psi_0$ by  holomorphic polynomials. In a radial gauge suitable for a central-symmetric confining potential \emph{coherent states}  of this space are 
\begin{equation}\la{W}\psi_{_V}= { Z^{-\frac 12}_{_V}}
e^{\frac{1}{\hbar}\sum_i^{N} V(z_i)}\psi_0,\;\;
 \psi_0=\Delta^{\beta}e^{-\sum_i^{N}\frac{|z_{i}|^2}{4\ell_B^2}},
 \end{equation}where \(\Delta\!\!=\!\prod^{N}_{i>j
}(z_i-\!z_j)\) and \(Z_{_V}\) is a normalization. 

A complex potential $V(z)$
is   analytic at infinity and such that \(4\pi \sigma=-\Delta V\) is real ($\rm{Im}\,\sigma=0$).
A  meaning of \(\sigma\) is a density of  "holes". From hydrodynamics perspectives they are  vortices, cf. \eq{HH}. 
 Also a set of permissible  operators is spanned by a product of holomorphic and anti-holomorphic operators.
 
 We denoted the averages of  symmetric operators in  a given \(V\)-state as
 \(\<{\cal O}\>_{_V}=\int \psi_{_V}^\ast
{\cal O}\psi_{_V}\prod_i \dd^2 r_i. \)

These states have been studied in \cite{WZ}. We mention an important  sum rule that follows from the  value of the dilatation operator \(-\frac 1N\sum_i\<(\vec r_i\cdot\vec\nabla_i)+(\vec r_i\cdot\vec\nabla_i)^\dag\>_{_V}=2\)
\begin{eqnarray}\la{sum}\frac 1N\sum_i\left(\frac{\<r_i^2\>_{_V}}{2\ell_B^2}-N\beta-\frac{1}{\hbar}\<\vec r_i\cdot\vec\nabla_i\rm{Re} V\>_{_V}\right)=1\!-\!\frac \beta 2.\end{eqnarray}

We construct the Hamiltonian based on a few defining properties: (i) Lauglin's 
w.f. is the ground state; (ii) the Hamiltonian density is an ordered  product of holomorphic and anti-holomorphic operators; (iii) long waves  of a FQH liquid are Galilean invariant; (iv) on  closed manifolds (no boundaries)  all  states are gapped \cite{F4}.  

Under these assumptions the zero Hamiltonian of the Chern-Simons theory is replaced by the Bernoulli energy
\begin{align}\label{B}
H=\int\frac { m_\nu}{2}\hat \vv^\dag\hat\rho\hat  {\vv}\,\dd^2 r,
\end{align}
where \(m_\nu=\frac{\pi\beta\hbar}{\kappa}\sim \frac{\hbar^2\ell_B^2}
{\Delta_\nu}\) is an effective mass obtained from the value of a gap,  $\hat\rho(r)=\sum_i\delta(r-r_i)$ is the density, and   \(\hat \vv=\hat v_x-i\hat v_y\) is the complex velocity. Velocity of  a particle is 
 \be\label{v}
  \frac{\ii}{2\hbar}m_{\nu}\hat\vv_i=\partial_{z_{i}}-\frac{e}{2c}{ A}(z_i)-
 \sum_{j\neq i}\frac{\beta }{z_{i}-z_{j}},\quad \beta=\frac{1}{\nu}.
 \ee
 Here \({ A}=A_x-iA_y\) is an external e.m. potential  $B=\nabla\times A$,  \(z_i=x_i+\ii  y_i\) is the complex  coordinate of a composite particle, \(2\p_z=\p_x-\ii\p_y\).  We occasionally use $\beta=1/\nu$. 
 
A subtle point of  this Hamiltonian is the definition of
the velocity operator \(\hat {\vv}\). It differs from the  velocity of individual electrons but corresponds to the velocity of "guiding centers"  or   "composite particles" - electrons with an  attached flux converting them to  bosons. The velocity of guiding centers  changes slowly in  long-wave excited states. This is the velocity  entering hydrodynamics.
We  emphasize  that the "mass velocity"   \eq{v} differs from the velocity obtained from electric currents carried either by
particles or  by guiding centers \cite{F1}. 

The Hamiltonian (\ref{B})  can be treated as a quantized version of 
 the "effective" Hamiltonians proposed and studied  in Refs.
\cite{GL}.

 \emph{4. Quantum hydrodynamics of the  FQHE liquid} describes dynamics of 
 velocity and density \emph{fields} when the number of particles is large. The fields are defined as  operators acting on averages  \(\<{\cal O}\>_{_V}.\) In particular, the velocity field is defined as   
\(\rho(r)\vv(r)\<{\cal O}\>_{_V}\!=\!\<\sum_j\delta( r-r_j)\vv_j {\cal O}\>_{_ V}\), where  \(\rho( r)=\<\hat \rho( r)\>_{_V}\).  In this representation the velocity
reads\be\la{p} m_\nu\vv=2\p_{z}(\pi_\rho
-\ii
V),\quad\pi_\rho=-\ii\hbar\frac {\delta}{\delta\rho}.\ee
In the restricted space  (\ref{W}) matrix elements of $\pi_\rho$
are imaginary, $\Delta V$ is real, hence the liquid is incompressible
\begin{align}\la{i}
\vec\nabla\cdot \vec\vv=0.\end{align} 
It is customary descryibe the incompressible flow in terms of the \emph{stream function}
\be\la{i1}
\vec\vv=\vec\nabla\times\Psi,\quad m_{\nu}\Psi=\ii\pi_\rho+
 {\rm Re} V.
\ee 
Eq. \eq{i1} gives an interpretation  to the deformation potential $V$. Its real part is the diagonal part of the stream function. Diagonal part  of vorticity and  energy  operators (\ref{B},\ref{p}) are  
 \begin{align}\la{HH}
 \<H\>_{_V}= \frac{1}{2m_\nu}\int|\nabla V|^2\rho
  \dd^2 r,\:\;m_\nu\<\vec\nabla\times\vec\vv\>_{_V}=-\Delta
V
\end{align}
Hydrodynamics of the FQHE fluid can be also seen as  dynamics 
 of the function \(V\).  
 
 \emph{5. Holomorphic fields, potential  incompressible flow  and Edge states.} In a system without  boundaries all  modes are gapped. If there is a boundary,  gapless  edge states emerge.  In order to focus on edge states it is sufficient to consider only a potential flow where a stream function is harmonic
\be\la{a} \vec\nabla\times
\vec \vv=0,\quad \Delta\Psi=0.\ee
Potential  flow corresponds to  deformations of the w.f. \eq{W} by analytic functions  inside the domain occupied by the liquid. All  singularities of \(V\) are outside of the domain (analytic functions do not exist on  closed manifolds, so as gapless modes). In this case  a FQH flow is  
 potential and  It has been shown in \cite{WZ} that a holomorphic deformation of   Laughlin's  state changes only the shape of the droplet,  leaving the  density  and the  area  unchanged. In the leading \(1/N \) order, the bulk density is uniform   \(\bar\rho={\nu B}/{\Phi_0}$. In a central-symmetric potential the droplet in the ground state is a disk with a radius $R=\sqrt{N/\pi\bar\rho}$. 

Incompressible potential flow with a free boundary and a constant density
is a standard subject in classical hydrodynamics \cite{Whitham}. Its extension
to the quantum case is straightforward.    
 Use the Green formula   and the Cauchy-Riemann
condition to  express the Hamiltionian
only through the boundary value of the fluid potential\be\la{BB}
 H=\frac{m_\nu\bar\rho}{2}\oint\Psi\p_n\Psi ds,\quad \<H\>_{_V}=
 \frac{\bar\rho}{2\ii m_\nu}\oint\bar V\dd {\rm } V.\ee The bulk Hamiltonian
vanishes.

 The  remaining task is to connect
the potential \(V\), or the velocity to the boundary  elevation $y(x)$.
Then  the  only governing equation is  the \emph{kinematic boundary condition} \cite{Whitham}  \begin{align}
\la{kbc}\dot y+v_x\nabla_x y+v_y=0.
\end{align}
  Later in this paper we show that 
\be\la{t1} v_x =c_0-\kappa\bar\rho y(x),\quad v_{y}=\kappa\eta\, \cdot y_{xx}^{H}\ee
where \(v_x, \,v_y\) are velocities of the inner layer tangential and normal to the unperturbed boundary.
Combing (\ref{kbc},\ref{t1}) and using $f(x)=\bar\rho y(x)$ we obtain the qBO \eq{BO}.

\emph{6. Chiral constraint} is a relation \eq{t1} between the velocity and the shape of the boundary of the droplet, hence a relation between $V$ and $y(x)$ in a state \eq{W}. This  relation    has been studied in \cite{WZ}. Here we obtain the constraint invoking Dyson's arguments \cite{D} and refine the results of \cite{WZ}. Dyson's arguments are  somewhat heuristic but transparent and short.  They are  somewhat heuristic but transparent and short. 

Let us express an expectation value \(\<{\cal
O}\>_{_V}\) as a path integral over the density field
$$\<{\cal
O}\>_{_V}=\int {\cal O}[\hat \rho] e^{-\beta F_{_V}[\hat \rho]}{\cal
D}\hat \rho.$$ The chiral constraint (aka \emph{loop equation} \cite{WZ}) is the saddle point condition ensured by a large number of particles 
\be\la{S}\left(\frac{\delta}{\delta\rho}-\beta \frac{\delta}{\delta\rho}F_{_V}[\rho]\right){\cal O}[\rho]=0.\ee 
The functional \(F_{{_{V}}}[\rho]\)
 can be treated as
 the free energy  of 2D-Coulomb plasma. It consists of energy and entropy.
\be
-\beta F_{{_{V}}}=\log|\psi_V|^2-\int \rho\log\rho\dd^{2} r.
\ee
 The entropy
 is the Jacobian of passing from  integration over particle coordinates to
a path integral over the density field.   

 In order to find the energy of the plasma  we write \(\sum_{i,j\neq
i}\log|r_i-r_j|=\sum_{i, j}\log|r_i-r_j+\ell\delta_{ij}|-\sum_i\log\ell(r_i) \), where  $\ell(r)$
is the mean  distance between
particles. Exclusion of "self-interaction" allows to replace sums  by integrals:
\(\sum_{j\neq i}\log|r_i-r_j|=-\frac 12\varphi(r_{i})-\log\ell(r_{i})\), where
 \be\label{CC}\varphi(r)=-2\int\log|r-r'|
 \rho\,\dd r^2\ee is the Coulomb potential of the plasma. This gives
 \be\beta F_{_V}=\int [\frac \beta 2(\varphi-\bar\varphi)+\beta\log\ell+\log\rho-
  \frac 2\hbar{\rm Re} V]\rho \dd^2 r,\la{F}\ee 
where \(\bar\varphi=-2\int_0^R\log|r-r'|\bar
 \rho \dd^2 r'=\pi\bar\rho r^2\) is the potential of a neutralizing
uniform background charge $\bar\rho$.

 A subtle point of this approach is the value of the mean distance between particles.
It is different in the bulk and close to the boundary. In the bulk the mean
distance is isotropic and \(\ell\sim 1/\sqrt\rho\). The short distance term and
the entropy term in \eq{F} sum up to \((\frac{\beta}{2}-1)\log\rho\).

Summing up we write the condition \eq{S} as   
\begin{align}
\text{bulk:}\quad2\pi \kappa^{-1}\Psi=
\varphi-\bar\varphi-(1-2\nu)\log \rho.\quad \
 \label{b}\end{align}

Close to the boundary, $\ell$ entering \eq{F} is the distance in the direction  normal to the boundary.
Since the mean distance along the boundary is  constant $\sim \ell_{B}$ we
have $\ell_B\ell\sim 1/\rho$. In this case the short distance term and
the entropy term sum up to \((\beta-1)\log\rho\).  

 Unfortunately,
Dyson's arguments miss some exponential corrections  which become important at the boundary. Notice that in the  case of the IQHE when  $\beta=1$ the term  \((\beta-1)\log\rho\)  vanishes.  The Dyson's arguments give $2\pi\kappa^{-1}\Psi=
\varphi-\bar\varphi$   . This equation treats the density as a step-function. Instead, the exact density of the IQHE is  \be\rho_I=\frac 12\bar\rho\;
 {\rm erfc}(\frac{y}{\sqrt{2}\ell_B})\approx
 \bar\rho\Big(\Theta(-y)+\frac{\ell_B}{y\sqrt{2\pi}}
 e^{-\frac{y^2}{2\ell_B^2}}\dots\Big),\nonumber\ee
   where \((-y)\) is the distance to the boundary. 
   
       This failure can be "repaired" by replacing   the plasma Coulomb potential \(\bar\varphi=\pi\bar\rho r^2\) of a neutralizing uniform charge $\bar\rho$    in \eq{F} by  the potential of the charge \(\rho_I\):     $\varphi_I=-2\int\log|r-r'|
 \rho_I(r')\dd^{2} r'\approx\pi\bar\rho r^2(1+
O(e^{-{y^2}/{2\ell_B^2}}))\). Then \begin{align}
\text{boundary:}\quad 2\pi\kappa^{-1}\Psi=
\varphi-\varphi_{I}-2(1-\nu)\log \rho.\;\;
 \label{bb}\end{align}
  This {\it ad hoc} procedure reflects a disreteness of particles. 
  
 So far we did not assume that the flow is potential. If the flow is potential  the Laplace operator 
 nulls the l.h.s. of \eq{b}. We obtain  a Liouville-type equation
 \be \la{L1}\text{boundary:}\quad\rho-\rho_I+\eta\Delta\log\rho=0.\ee We do not have a satisfactory  mathematical justification of this equation, but it  treats correctly exponential corrections  becoming important at the boundary. Also, numerical solution of the Liouville equation \eq{L1} appears to be undistinguishable from a numerical \emph{ab initio} simulation of the Laughlin's wave function \eq{W} \cite{Abanov,Shytov}.  

Confining potential is easy to incorporate into (\ref{bb}).  It shifts the current \(\rho\vec\vv \)  by the Hall current $\frac{\nu}{h}\vec\nabla\times U$. Taking the curl of \eq{bb} 
and multiplying by $\rho$
we obtain \cite{F2}
\begin{align}\label{PP}
\frac{2\pi}{\kappa}(\rho\vec\vv\!-\!\frac{\nu}{h}\vec\nabla\!\times\! U)
=\rho\vec\nabla\!\times\!(\varphi\!-\!\varphi_I)\!
-2(1\!-\!\nu)\nabla\!\times\!\rho
\end{align}
We assume that
 \(\ell_B^{2}\nabla^2_yU\ll\Delta_\nu\).  Then the  effect  of the confining potential at the inner   boundary is reduced
to a shift of the velocity along the boundary $v_x\to v_x-c_0$.
Eqs. (\ref{b}-\ref{PP}) can be checked against the sum rule \eq{sum}. Using \eq{p} and an exact formula  $\int  r^2\rho_{I}\dd^2 r=\frac{R^{2}}{2}(N+1)$.  
we can write the sum rule as \be\la{sum1} \frac{1}{\kappa N}\int\dd^2 r[\vec r\times(\rho\vec\vv)]\!-\frac{1}{\pi R^2}\int  r^2(\rho-\rho_{I})\dd^2
r=4\eta. \ee  
\emph{7. Boundary double layer and its dipole moment}  Eq.(\ref{L1}) has a profound consequence. It shows that  the density behavior is singular: on approaching the boundary the density oscillates and shoots up before falling down.  The overshoot has been observed numerically in \cite{overshoot} and also discussed in \cite{WZ}. A detailed  structure of the overshoot is complicated and not  yet  satisfactory understood. However, we know that in a rough but sufficient  approximation  it is a double layer as is in Fig. \ref{dipole}. As we see in a moment, only the Coulomb potential of the  layer  - the boundary dipole moment  (per unit length)  enters the edge dynamics. 

The value of the dipole moment follows from the  sum rule (\ref{sum},\ref{sum1}) evaluated sufficiently
close to the boundary where normal velocity in \eq{sum1} can be neglected.
 It is\begin{align}\la{d}
\eta=\int y(\rho-\rho_I)\dd y=\frac{1-\nu}{4\pi}.
\end{align}   
A detailed structure of the boundary layer follows from \eq{L1}. Iterations of this equation  
  allow us to  conjecture the asymptotic expansion for the density from  the bulk side. In units of $\ell_{B}$ it reads
  \be\nu \rho = \bar\rho\Big(\nu+
 \frac{e^{-{\xi^2}/{2}}}{\sqrt{2\pi}}(2\pi\eta\,\xi+O(\xi^{-1})\Big),\quad \xi<0.
 \la{e5}\ee
 At \(\ell_B\to 0\) the leading term of \eq{e5} is a double layer \cite{Abanov}
\begin{align}\la{A}
\rho(y)\approx\rho_I(y)+\eta\delta'(y).
\end{align}  
The following arguments are based on this major formula.

\emph{8. Transformation of velocities} Now we are in a position to compute  the velocity  in terms of the boundary elevation $y(x)$ and to complete  the governing equation  \eq{kbc}. As in \eq{kbc} we assume the density moves together with the boundary  $\rho(x,y)=\rho_0(y-y(x)),$ where $\rho_0(y)$ is the density of  the ground state and compute the boundary value of the  Coulomb potential $\varphi$ \eq{CC} entering the chiral constraint \eq{bb}.
 This exercise is equivalent to the Hadamard formula for a variation of solution of a boundary value problem upon a variation of the boundary. Computing $\int\log|r-r'|\rho_0(y'-y(x'))dx'dy'$, we shift the variable $y'\to y'+y(x')$, subtract $\rho_I$ and expand  in $y(x)$. We obtain \(\varphi(x,y)-\varphi_0(y-y(x))\approx\)$$ 2\int dx'(y(x')-y(x))\int\frac{y-y'}{|r-r'|^2}[\rho_0(y')-\rho_I]dy'-2\pi\bar\rho y y(x).$$ The integral over $y'$ is localized inside the boundary layer. If we choose $y$ to be on the inner boundary of the layer
the range of $|y-y'|\sim\ell_B$.  We can safely replace $|r-r'|^{-2}$
in the integral
by $(x-x)^{-2}$ and obtain the transformation law of the stream function and the velocity under a displacement of the boundary  by $y(x)$ (valid for any shape of the boundary)
\begin{align}
& \Psi(x,y) = \Psi _{0}(y-y(x))-\kappa\left( \bar\rho y\cdot y(x)+\eta\cdot y_{x}^{H}\right), \\
& v_x =v_{0x}-\kappa\bar\rho y(x),\quad v_{y}=v_{0y} +\kappa\eta\, \cdot y_{xx}^{H}.
\la{t}\end{align}
 For a flat boundary  $v_{0x}=c_0,\;v_{0y}=0$.  

A relation between $V$ and $y(x)$ follows from  \eq{t}
\be\la{V}
(\pi\beta \hbar)^{-1}V'= \frac{ \bar\rho}{ 4\pi}\int \frac{ y(x')dx'}{z-x'}+
\frac{\eta}{2\pi}\int\frac{y(x')dx'}{(z-x')^3}.\ee
\\

The Benjamin-Ono equation \eq{BO} and the current algebra \eq{f}
follow from the transformation law \eq{t},  and the value of the dipole moment \eq{d}. Alternatively the  Benjamin-Ono equation  can be  also obtained from  the boundary  Hamiltonian \eq{BB} expressing it through $f$ with the help of \eq{V}.

The author acknowledges help by A. Abanov, E. Bettelheim, A. Zabrodin and T. Can on many aspects of this work, useful discussions with W. Kang, I. Gruzberg, A. Cappelli and B. Spivak and thanks A. Shytov for sharing numerical data. The works was supported by NSF DMR-0906427, MRSEC under DMR- 0820054.
 

\begin{thebibliography}{10}
\bibitem{F1}A difference between the electric current and \eq{v} attributes to  a term 
\(\frac{e}{2m_\nu c}\int B \hat \rho\, \dd^2 r \) omitted in the Hamiltonian  \eq{B}
 This term is not important for density waves  if magnetic field is 
  uniform.
\bibitem{F2} A redefinition  of  the current \(\rho\vec \vv\to
\rho\vec\vv+\xi\vec \nabla\times\rho\) changes the vorticity $\vec\nabla\times \vec\vv$ by \(\xi\Delta\log\rho,\)
 and, therefore, the coefficient  in front of \(\log\rho \) in (\ref{b},\ref{bb}),
without affecting divergency of  velocity.
 A choice of the velocity depends on a physical setting \cite{F1}, but this ambiguity
does not affect  density waves on the edge.
\bibitem{F3} Contrary to the asymptotic expansion for $\rho_{I}$ eq. \eq{e5} is valid only for $\xi<0$. Asymptotic expansion for $\xi>0$:  $\rho\sim\ e^{-\xi^{2}/2} \xi^{-\beta}\Big(1+O(\xi^{-1})\Big)$ 
This  asymptote will be affected by the confining potential.
\cite{WZ1}.
\end{thebibliography}
\end{document}